\documentclass[aps,prl,floats,twocolumn]{revtex4}

\usepackage{graphicx}
\usepackage{docs}

\def\be{\begin{equation}}
\def\ee{\end{equation}}
\def\ba#1{\begin{array}{#1}}
\def\ea{\end{array}}
\def\bn{\begin{enumerate}}
\def\en{\end{enumerate}}

\def\r{\right}
\def\l{\left}

\def\at{a_{\tau}}

\def\xt{_{(x,\,\tau)}}
\def\scs{SC$^{\star}$~}

\def\summ{\sum\limits}

\def\ax{a_x}
\def\vms{c}

\def\j{K}
\def\xz{\lambda_Q}

\def\aeff{\alpha}
\def\o{\omega}
\def\lb#1{\label{#1}}

\begin{document}

\title{Superconductor-to-normal transition in finite nanowires}
\author{{\sc Gil Refael$^1$, Eugene Demler$^2$, Yuval Oreg$^3$}\\
{\small \em $^1$Dept. of Physics, California Institute of Technology, MC 114-36, Pasadena, CA 91125\\
$^2$Dept.\ of Physics, Harvard University, 17 Oxford St., Cambridge MA, 02138\\
$^3$Dept. of Condensed Matter, Weizmann Institute of Science, Rehovot, 76100, Israel}}

% -------------------------%
\begin{abstract}
In this paper we discuss the interplay of quantum fluctuations and
dissipation in uniform superconducting nanowires. We consider a
phenomenological model with superconducting and normal components, and
a finite equilibration rate between these two-fluids. We find that 
phase-slip dipoles proliferate in the wire, and decouple the
two-fluids within its bulk. This implies that the the normal fluid
only couples to the superconductor fluid through the leads at the
edges of the wire, and the {\it local} dissipation is unimportant.
Therefore, while long wires have a superconductor-metal transition tuned by
local properties of the superconducting fluid, short wires have a transition when
the {\it total} resistance is $R_{\rm total}=R_Q=h/4e^2$. 
\end{abstract}
\pacs{PACS Numbers:}

\maketitle
%\maketitle

% -----------------------general introduction%

%TWO FLUID REFERENCES

Quantum phase transitions have long been at the forefront of condensed
matter theory. Especially interesting are systems of reduced
dimensionality and size, where fluctuations are enhanced, and ordering is illusive, and far from being
expalined by mean field theory. Such systems exhibit a surprising degree of
universality; for instance, as observed in Refs. \cite{paalanen,
  Haviland2B}, a mesoscopic
Josephson junction shunted by a resistor $R$ undergoes a
(so-called Schmid) transition between a Coulomb-blockade (normal) and superconducting phase when
the shunt resistor is $R=R_Q=h/4e^2=6.45k\Omega$ \cite{Schmid,Chakravarty,bulga}. Fluctuations of the
superconducting phase angle, i.e., phase-slips, induce this transition;
they also control the onset of superconductivity in long thin
wires and Josephson junction chains \cite{Dynes0,Haviland2A,
  japan, arutyunov2, chang1}, where the competition between local charging energy,
which creates phase-slips, and the
superconducting stiffness tunes the transition \cite{ZAIKIN1997,KorshunovA,Bobbert}.

\begin{figure}
\includegraphics[width=6.5cm]{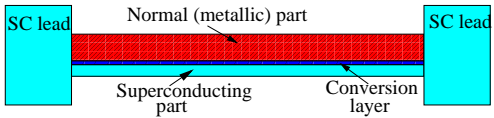}
\caption{
A two-fluid model of a superconducting nanowire with a normal part,
depicted as a separate region. In
  order for electrons to change from normal to superconducting, they
  need to pass through the conversion layer, which has a finite
  conductivity. Proliferated phase-slip dipoles inhibit two-fluid
relaxation, and make the conversion layer insulating, which renders
the normal part as an effective single shunt resistor. \label{fig1}}
\end{figure}

Our focus is experiments on $Mo_{79}Ge_{21}$ (amorphous) nanowires as narrow as 5nm-15nm. Resistance vs.
temperature curves showed a transition between superconducting
(resistance decreasing upon cooling) and normal, or weakly insulating,
(resistance non-decreasing upon cooling) behavior. A first set of
measurements on wires of various diameters, and lengths
$100nm<L<200nm$, showed a remarkable result: a transition 
when the {\it total} resistance of the wire was
$R_Q=h/4e^2=6.45k\Omega$ \cite{Bezryadin}, as if the entire wire was a single
shunted junction. But the coherence length of $MoGe$ is $\xi< 10nm$, 
thus the wire should differ dramatically from a single junction.     
Indeed, later experiments on longer wires, $200nm<L<1000nm$, showed a weak transition
that depended on the {\it resistance per length} or {\it
  cross-section} of the wires, i.e., on a local quantity, rather
than the total resistance \cite{Tinkham}. Later experiments
\cite{BollingerUP}, could neither prove nor disprove the global
nature of the transition in the shorter wires. 

In this paper, we describe nanowires using a two-fluid model, which
assumes that Cooper pairs couple to a normal electron fluid, which
provides local dissipation (Fig. \ref{fig1}). Remarkably, we
find that {\it at sufficiently low temperature, the normal and superconducting fluids
within a continuous nanowire decouple due to quantum phase
fluctuations}, thus rendering the local dissipation unimportant. Therefore, the superconducting degrees of freedom can
only couple to the dissipative normal fluid at the leads, on the edges
of the wire, where they couple to its {\it
  total} normal-state resistance, $R_{\rm total}$. As a result, we show
that indeed short wires may undergo a global
dissipative Schmid transition tuned by the total wire
resistance, when $R_{\rm total}=R_Q$. By short, we mean wires with
length $L>\xi$, but shorter than both the thermal length, $L<\hbar \vms/T$
(with $\vms$ the Mooij-Sch\"on velocity, and $T$ being the lowest
temperature in the experiment)\cite{footnoteT}, and the
'quantum length' $A_C R_Q/\rho$, (with $\rho$ the specific resistance,
and $A_C$ the largest cross-section area where quantum phase slips are
not competely suppressed)\cite{footnoteA}. After our work was completed,
  this result was verified in $MoGe$ wires with $50nm<L<300nm$ \cite{Bezryadin-TBP}. Below we derive the two-fluid model, show how phase-slip
  dipoles decouple the normal and superconducting fluids, and apply to
  model to the case of a finite nanowire.

\begin{figure}
\includegraphics[width=9.5cm]{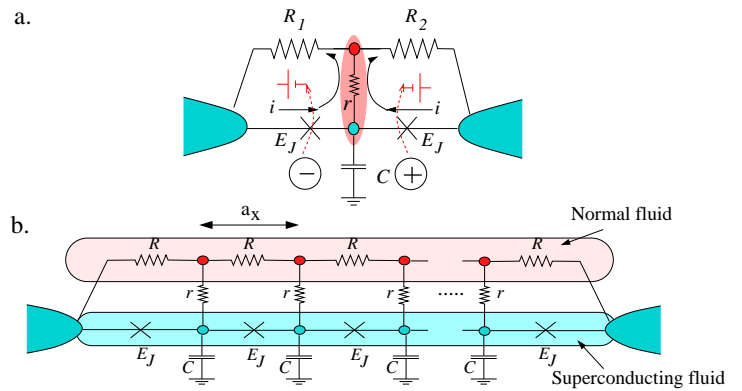}
\caption{(a) Single two-fluid grain (pink ellipse) between two leads: dipoles produce two opposing
  voltage spikes on the two junctions, which oppose super-to-normal conversion currents leaving the superconducting part 
  of the grain (bottom circle in ellipse) and entering the normal part
  (top circle). When dipoles proliferate, the
 superfluid-normal conversion resistance (and time) effectively diverges,
  $r\rightarrow\infty$, and the normal and superfluid are completely
  decoupled. (b) We begin our study with a chain of two fluid grains,
  which is a discretized version of the nanowire in Fig. 
\ref{fig1}. Note that $1/r=\Upsilon a_x$ and $R=a_x \rho$.  
 \label{fig2}}
\end{figure}

The hint of a Schmid transition, the long
resistive tails seen in experiments, and the strong disorder of the
$MoGe$ nanowires suggest the presence of local dissipation, which
motivates the two-fluid model approach. We assume that charge can flow in the
nanowires in two ways: as diffusive normal electrons with resistivity
$\rho$ - normal fluid - and as bosonic Cooper pairs - superfluid. The
normal fluid stems from 
strong disorder and phase-fluctuations, which 
suppress the proximity effect and possibly give rise to normal regions and a finite density of states
for single electrons at the Fermi level.
The two fluids can have a different chemical
potential, and can exchange charge with a finite, bare, relaxation
time, $\tau_r=\Upsilon^{-1}$, in a
bulk system (see Fig. \ref{fig1}, and Fig. \ref{fig2}b for a discrete model). This is related to the branch imbalance
relaxation time \cite{RiegerScalapino, ImbalanceReview}, first measured by Clarke \cite{ClarkeImbalance} in Sn
wires.  

Before plunging to the analysis, note that earlier works on similiar
models considered only the perfect normal-super fluid coupling, $\Upsilon=\infty$ case, and found a superconducing-metal transition
tuned by the resistance per length \cite{ZAIKIN1997,KorshunovA,
  Bobbert,Golubev-Zaikin}, as did
Refs. \cite{Yuval1,Yuval2}. Alternative approaches assumed external
dissipation coupled to the leads but not to the bulk of the wire \cite{Buechler}, or discussed the onset of
superconducting correlations and neglected phase fluctuations
\cite{SachdevTroyerWerner}.  

Indeed, in our model as well, sufficiently long but finite wires should exhibit a
SC-normal crossover tuned by their cross-section area, which sets the bare
fugacity of quantum phase slips \cite{footnoteA}, as well as their stiffness. But quantum fluctuations in the form of
 phase-slip dipoles, make the Cooper-pair to normal-electron conversion
 rate vanish at $T=0$ in the bulk of the nanowire:
 $\Upsilon\rightarrow 0$. As
 claimed above, this leads to a true Schmid transition for short wires,
 which effecively become a short dissipationless superconducting wire,
 shunted through the leads by the total normal-state resistance $R_T$.

The crucial two-fluid decoupling is already evident in a simple two-junction
system (Fig. \ref{fig2}a) \cite{Refael2003, WernerRefael}. When $r=0$
(i.e. vanishing conversion resistance), the two
junctions in the system are independent in the
d.c. limit. Phase slips - events where the phase across a Josephson
junction tunnels by $2\pi$ - create a sudden voltage drop that opposes
any supercurrent flowing, and thus induce dissipation. A Schmid transition occurs in each junction when
$R_i=R_Q$ ($i=1,2$). When $r>0$, the two junctions become coupled, and
phase-slips may form bound dipoles: simultaneous phase-slip and anti-phase-slip in
the two junctions. Remarkably, dipoles do not destroy the coherence between
the two leads, since they produce equal and opposite voltage drops. Nevertheless, as single phase slips block
supercurrents across their Josephson junctions when they proliferate,
dipoles block the normal-superfluid conversion channel: a conversion current $2i$ (Fig. \ref{fig2}a) flowing across
$r$, with no lead-to-lead current, implies a current $i$ on both
junctions, but in opposite directions. $i$ couples directly to the voltage drop of
the dipoles; when proliferated, they block this current mode, and thus {\it decouple the normal
  and super fluids}. Phase slip dipoles
proliferate roughly when $r>R_Q$.  In this case a global Schmid
transition takes place when $R_1+R_2=R_Q$.

%\begin{figure}
%\includegraphics[width=9.5cm]{fig25}
%\caption{We first study a chain of two fluid grains,
%  which is a discretized version of the nanowire in Fig. 
%\ref{fig1}. Note that $1/r=\Upsilon a_x$ and $R=a_x \rho$.  
% \label{fig25}}
%\end{figure} 

%{\bf the next system to consider: JJarray}

Next, we generalize the normal-super fluids decoupling to wires, first
using a discrete model (Fig. \ref{fig1}b), and then taking its
continuum limit. Starting with an infinite chain of mesoscopic two-fluids grains
(Fig. \ref{fig1}b) \cite{JJarray}, the low-energy 
action for the chain is given in terms of a 2d gas of phase-slips, with
interaction:
\be
p_1p_2 \l(\j \log\frac{\at}{\sqrt{x^2/\vms^2+\tau^2}}
+\aeff e^{-|x|/\xz}
\log\frac{\at}{|\tau|}\r).
\lb{inter}
\ee
$\vms=\ax \sqrt{E_J E_C}/\hbar$ is the Mooij-Sch\"on velocity
\cite{Mooij}, and $\at=\ax/\vms$. $p_i=\pm$ is the phase slip polarity. The first term is the usual isotropic interaction of a 1+1 XY
model due to the plasmons in the Josephson junction array;
$\j=2\pi\sqrt{E_J/E_C}$, $E_C=(2e)^2/C$. The second term is due to the
dissipative interaction: $\aeff=max\{\frac{R_Q}{\sqrt{rR}},\frac{R_Q}{R}\}$, and
$\xz=max\{\ax\sqrt{r/R}=1/\sqrt{\Upsilon\rho},\ax\}$ is a new length-scale that arises from
the two-fluid finite relaxation time. As in the two-junction case,
dipoles must be explicitly included in the low-energy description of
this model  \cite{JJarray,
  KorshunovA, Bobbert}. We denote the fugacity of
single phase slips as $\zeta$, and the fugacity of a dipole with
moment $n$ as $\eta_n$. For completeness, we quote here the explicit field theory
for the infinite chain: 
\be
\ba{l}
\int \frac{d\o dk}{(2\pi)^2}\l[\l(\vms
  k^2+\frac{1}{\vms}\o^2\r)\frac{\theta_{(k,\o)}^2}{4\pi\j}+\frac{r}{4\pi R_Q}
  |\omega|\l(k^2+\frac{R}{r}\r)\psi_{(k,\o)}^2\r]\vspace{2mm}\\
-\int d\tau \summ_i\l[\zeta \cos (\theta_i + \psi_i)+\eta_n\cos(\Delta_n\theta_i+\Delta_n\psi_i)\r]
\label{action}
\ea
\ee
with $\vms=\ax \sqrt{E_J E_C}/\hbar$, and $\theta,\,\psi$ mediating the plasmon and
dissipative interactions, respectively. At high energies  $\eta_n\sim
\zeta^2$, and $\Delta_n f_i=f_{i+n}-f_i$.  This is a
representation dual to the SC phase representation, hence
$\exp(i\psi_i+i\theta_i)$ is the operator that creates a phase slip on
junction $i$. 

It is useful to compare the relatively complicated interaction between
phase slips in an infinite chain, with that of phase slips in a single
Josephson junction. In a single junction the interaction is:
$p_1p_2 \frac{R_Q}{R} \log\frac{\at}{|\tau|}$.
The Schmid transition, which marks phase-slip proliferation, occurs when the gain in entropy due to
separating a phase-slip from an anti phase slip, $S=\log (\at T)$
equals the required interaction energy, $\frac{R_Q}{R} \log(\at
T)$.  Employing the same argument for the two-fluid Josephson chain
yields the approximate SC-normal phase boundary:
$\j+\frac{1}{2}\aeff\sim 4$.
This transition is essentially the 1+1 Kosterlitz-Thouless-Berezinski (KTB)
transition of the Josephson junction array in accordance with Ref. \cite{ZAIKIN1997}
(for a more refined analysis see Ref. \cite{JJarray}). 
But this argument, as well as the interactions in
(\ref{inter}), ignores phase-slip dipoles. When dipoles are
proliferated, the normal- and superfluids decouple, and the
superconducting part of the wire exhibits a
SC-normal KTB transition when $\j\sim 4$. {\it Phase-slip dipoles}, we
find, proliferate when:
\be
\frac{2R_Q}{R}(1-e^{-\ax/\xz})<1.
\label{dipolep}
\ee
The left-hand side is the strength of dipole interaction, which consists of the
self interaction of the slip and anti slip and also their mutual
interaction. $\ax$ is the distance between grains in the model.

\begin{figure}
\includegraphics[width=7.5cm]{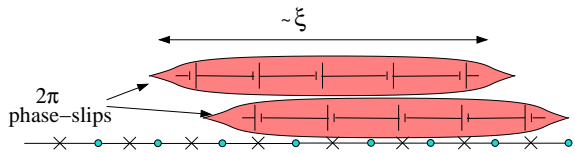}
\caption{To describe a continuous wire, we take the limit
  $\ax\rightarrow 0$ while keeping the coherence length and size of
  phase slip, $\xi$, fixed. This implies that phase slips are
  spread over $\xi/\ax$ junctions, and can form dipoles with
  separation $x<\xi$. The voltage signs symbolize the voltage drop
  caused by a phase slip.
\lb{fig3}}
\end{figure}

In the continuum limit, {\it dipoles always proliferate} and cut off the
superfluid-normal conversion. The continuum limit implies $\ax\rightarrow 0$; but this makes a Josephson junction
(and therefore also a phase slip on a junction) shrink to length zero. But
a phase slip occurring on physical nanowires has a characteristic length
$\xi$ (coherence length). To reconcile this we allow
phase slips to smoothly spread over $\sim\xi/\ax$
junctions \cite{JJarray}. Technically, we transform the zeta term in
Eq. (\ref{action}) as $\cos(\psi+\phi)_{\xt}\rightarrow
\cos\frac{\xi}{\ax}\sum_r f(r)(\psi+\phi)_{(x+r,\,\tau)}$, where
$f(r)$ is a smooth, normalized, function centered around $0$, with
width $\xi$. We similarly treat the dipole $\eta_n$ terms. The smearing
reflects that in nanowires, phase slips can have an almost arbitrary
overlap, $\Delta<\xi$ (Fig. \ref{fig3}), with other phase slips. The
continuum generalization of Eq. (\ref{dipolep}) is that dipoles
proliferate when: 
$\frac{R_Q}{R_{\xi}}\l(\frac{\Delta}{{\rm max}\{\xz,\,\xi\}}\r)^2<1$
(where $R_{\xi}=R_{\rm total}\xi/L=\rho\xi$). Thus
for any $R_{\xi},\, r$ there is a separation $\Delta_c$ below
which dipoles proliferate. By incorporating the above analysis and the
appropriate screening terms in Eq. (\ref{action}) we see that the normal-superfluid conversion is cut off at temperatures
(or frequencies) $T\sim \zeta_0^2\at$, where $\zeta_0$ is the
bare fugacity of phase slips, and $\at^{-1}\sim (\xi/c)^{-1}$ is the
UV cutoff in (\ref{action}).

\begin{figure}
\includegraphics[width=9cm]{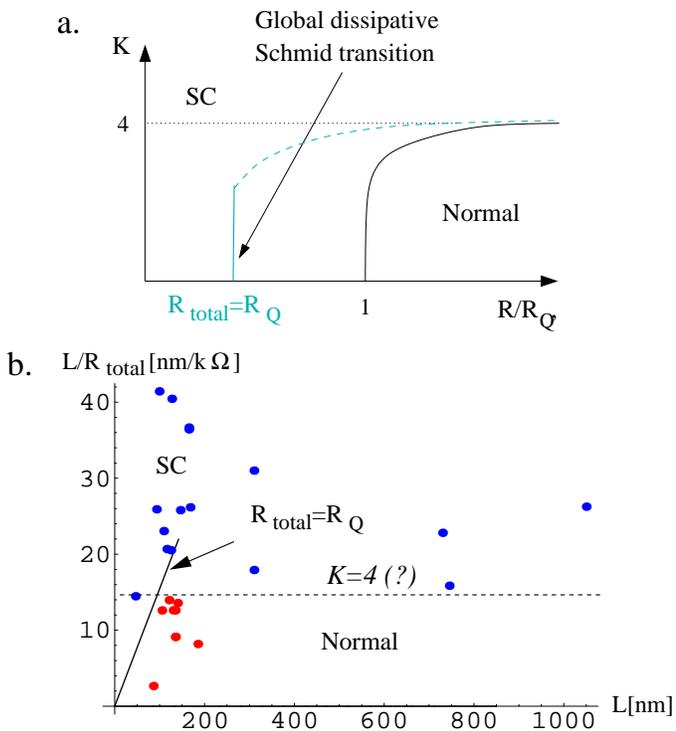}
\caption{(a) Phase diagram of the two-fluid chain and wire. The
  black line describes (roughly) the SC-normal boundary of an
  infinite two-fluid chain (Fig. \ref{fig2}b), with $r<R_Q$;
  both $K=2\pi\sqrt{E_J/E_C}$ and $R$ are local quantities. When
  $r>R_Q$, normal-superfluid conversion is cut off, the local
  dissipation becomes unimportant, and only the  horizontal line
  applies. In finite continuous wires (grey line), the transition
  takes a dissipative nature when the wire is short, and occurs when
  $R_{\rm total}=R_Q$; longer wires have a local crossover - dashed line - tuned by the superconducting stiffness
  $\sim\j$. (b) Phase diagram for $MoGe$ nanowires of
  Refs. \cite{Bezryadin,Tinkham, BollingerUP}. The y-axis,
  $L/R_{\rm total}\propto A/\rho_{MoGe}$ is proportional to the surface
  area, which is proportional to $\j$. The x-axis is the length. The blue dots are
  insulating samples, whereas the red dots are superconducting. The
  horizontal dashed line marks the long-wires cross over, while the
  diagonal black line marks the transition line $R_{total}=R_Q$.
\label{fig45}}
\end{figure}

Thus a finite continuous wire is effectively described by a  
chain of Josephson junctions, shunted by the {\it global resistance}
in the chain. Phase slips now exhibit an interaction due to the plasma
waves in the chain, $\j \log\frac{\at}{\sqrt{x^2/\vms^2+\tau^2}}$, and
also due to the dissipation through the normal resistance, which is
couple through the leads: $\frac{R_Q}{R_{\rm total}}\log
\frac{\at}{|\tau|}$. Naively, a transition will now occur when:
\be
\j+\frac{R_Q}{R_{\rm total}}\sim 4
\ee
(see Fig. \ref{fig45}a). Less naive considerations show that the KTB transition, tuned by
$\j$ is a cross-over for lengths $L<\vms/\hbar T$, and an
even stronger effect may appear due to the
bare fugacity of phase slips being exponentially suppressed
with $\j$. The total resistance, however, still drives a Schmid transition when
$R_{\rm total}=R_Q$. 

Our conclusions could be easily related to the nanowire experiments
\cite{Bezryadin, Tinkham, BollingerUP, Rogachev}. 
In long wires, we expect a SC-Normal crossover tuned by
stiffness (as in \cite{ZAIKIN1997,Golubev-Zaikin}), but in short
  wires, we expect a Schmid transition tuned by the total
normal-part resistance.  In Fig. \ref{fig45}b we recast the diagram of
Fig. \ref{fig45}a for the $MoGe$ nanowire experiments, plotting
$L/R_{\rm total}\propto A$ vs. $L$, with $L$ the length of the wire, and $A$ its
cross section area. The diagonal line marks $R_{\rm total}=R_Q$. Above it we expect $T=0$ superconductivity. The
horizontal line marks the SC-normal cross-over in longer
wires. This line most probably arises from the exponential dependence of the bare
quantum-phase-slip fugacity on thickness\cite{footnoteA}, but may
  also be associated with a KTB transition at
$\j\sim 4$, or a fermionic $T_C$ suppression mechanism,
  which also depends on $R_{\xi}/R_Q$ \cite{Yuval2}. After completing the analysis described here, Bezryadin and coworkers
measured a large number of short samples with $L<150nm$. These show near perfect fit with our
prediction of a universal transition at $R_{\rm total}=R_Q$ for shorter
wires \cite{BollingerUP, Rogachev} \footnote{Note that B\"uchler et al. \cite{Buechler}, considered a
SC-Metal transition tuned by the resistance in the measuring
circuit - this results in similar physical behavior as our model.}.

The application of our simple theory to the nanowire experiments
requires several caveats. (1) It is natural to associate
the resistance of the nanowire devices at temperatures just below the
SC transition of the leads, with the total normal-part nanowire resistance, $R_{\rm total}$. It is unclear,
however, how this resistance is related to the normal-state resistance
of the nanowires at temperatures above the bulk critical temperatures
for $MoGe$. (2) In addition, the origin and precise
nature of normal electrons in the wires is unknown. Possibly, phase
fluctuations or the strong disorder stifle the proximity
effect and give rise to a normal part. Particularly, if the normal-superconductor relaxation rate is
indeed suppressed, each phase slip gives rise to a long-lived population
of quasi-particles, as is the case in Ref. \cite{Dolan}
where dissipation at phase-slip centers is investigated. (3) The
resistance vs. temperature curves measured
by Bezryadin on the superconducting side show sharp exponential,
activated-like, decay of the resistance, contrary to a naive quantum
phase-slip theory, where an algebraic dependence of the resistance
on temperature is expected. Similarly, the wires remaining
normal show a weakly insulating behavior, with a charge-gap that
corresponds to the Coulomb-blockade of the leads
\cite{Rogachev}. These observations do not contradict the
possibility of a Schmid transition, and can be understood by also
considering the effective dissipation produced by a finite density
of phase slips, which is too large to justify the perturbative
analysis pursued here. Such considerations appeared to describe the intermediate-coupling regime of the
two-junction system \cite{WernerRefael}. Using the results presented
here regarding the Schmid transition, but adding a finite density of
phase-slips, we successfully explained the sharp temperature dependence of short
wires in Refs. \cite{Dganit1,Dganit2}. 

The main result of this paper is the divergence of the
normal-superfluid relaxation time, $\tau_r=\Upsilon^{-1}$, in continuous uniform nanowires due to quantum fluctuations. Apart
from the direct application of our theory to the $MoGe$ nanowire
experiments, this effect could be directly investigated in
meso and nanoscopic systems where quantum fluctuations are apparent at
relatively high temperatures. Some early experiments in this
direction on nanostructures not uniform enough, but with quantum fluctuations are described in Ref. \cite{Arutyunov}.
In future work we hope to address the issues of the origin and nature
of the normal-part in nanowires, and its interplay with phase-slip
density, and the diverging relaxation time.

We would like to especially thank
D.S. Fisher, who collaborated with us on much of the research reported
here. We also thank A. Bezryadin, A. Bollinger,
P. Goldbart, B.I. Halperin, D. Meidan, D. Podolsky, D. Shahar for many helpful discussions. G.R. thanks the Technion ITP and the Weizmann institute for their
generous hospitality. Y. O. Acknowledges support of the ISF. 

\bibliography{thebib}

\begin{thebibliography}{36}
\bibitem[{foo({\natexlab{a}})}]{footnoteT}
\bibinfo{howpublished}{The upper bound marks the crossover length where finite
  spatial-size corrections to the thermodynamic behavior occur. To get a sense
  for this scale we note that $L_{CO}=\hbar c/T\approx \xi
  \sqrt{N_{\perp}}\Delta_0/T$, where $\Delta_0$ is the BCS gap, and $N_{\perp}$
  is the number of transverse channels.}
\bibitem[{foo({\natexlab{b}})}]{footnoteA}
\bibinfo{howpublished}{The fugacity of quantum phase slips is $\zeta\sim e^{-c
  R_Q/R_{\xi}}=e^{-c A R_Q/\rho\xi}$, with $c$ of order 1, $R_{\xi}=R_{\rm
  total}\xi/L$ and $A$ the cross-section area \cite{Golubev-Zaikin}). In wires
  too thick, $A>A_C$, no quantum phase slips would be observed. $A_C$ depends
  on the prefactor of the exponent in $\zeta$, which could be quite large.}
\end{thebibliography}

\end{document}